\documentclass{isprs} 
\usepackage{subfigure}
\usepackage{setspace}
\usepackage{amssymb}
\usepackage{pifont}
\newcommand{\xmark}{\ding{55}}%
\usepackage{geometry} 
\usepackage{epstopdf}
\usepackage[labelsep=period]{caption}  
\usepackage[british]{babel} 
\usepackage{amsfonts}
\usepackage[hang]{footmisc}
\usepackage{lipsum} 
\usepackage{amsmath}
\usepackage{nicefrac}
\usepackage{booktabs}
\usepackage{float}
\usepackage{stfloats}
\usepackage{svg}
\usepackage{placeins}

\newcommand{\id}[2]{{\color{gray}#1}{\color{pink}#2}}

\newcommand{\df}{dual-form }

\usepackage[capitalise]{cleveref}
\crefname{figure}{Fig.}{Figs.}
\crefname{table}{Table}{Tables}
\crefname{equation}{Eq.}{Eqs.}
\begin{document}

\date{}

\title{Comparative analysis of dual-form networks for live land monitoring using multi-modal satellite image time series}
\author{Iris Dumeur\textsuperscript{1,2}, Jérémy Anger\textsuperscript{1,2}, Gabriele Facciolo\textsuperscript{2,3}}
\address{
	\textsuperscript{1 }Kayrros SAS\\
	\textsuperscript{2 }Universit\'e Paris-Saclay, ENS Paris-Saclay, CNRS, Centre Borelli, 91190, Gif-sur-Yvette, France\\
    \textsuperscript{3 }Institut Universitaire de France

}
\abstract{
Multi-modal Satellite Image Time Series (SITS) analysis faces significant computational challenges for live land monitoring applications. While Transformer architectures excel at capturing temporal dependencies and fusing multi-modal data, their quadratic computational complexity and the need to reprocess entire sequences for each new acquisition limit their deployment for regular, large-area monitoring. This paper studies various dual-form attention mechanisms for efficient multi-modal SITS analysis, that enable parallel training while supporting recurrent inference for incremental processing. 
We compare linear attention and retention mechanisms within a multi-modal spectro-temporal encoder. To address SITS-specific challenges of temporal irregularity and unalignment, we develop temporal adaptations of dual-form mechanisms that compute token distances based on actual acquisition dates rather than sequence indices.
Our approach is evaluated on two tasks using Sentinel-1 and Sentinel-2 data: multi-modal SITS forecasting as a proxy task, and real-world solar panel construction monitoring. Experimental results demonstrate that dual-form mechanisms achieve performance comparable to standard Transformers while enabling efficient recurrent inference. The multi-modal framework consistently outperforms mono-modal approaches across both tasks, demonstrating the effectiveness of dual mechanisms for sensor fusion. The results presented in this work open new opportunities for operational land monitoring systems requiring regular updates over large geographic areas.
}

\maketitle


\sloppy

\section{Introduction}
Decades of Earth observation programs, from Landsat to Sentinel, have produced frequent and consistent records of land surfaces. Organized as multi-modal Satellite Image Time Series (SITS), these data support a wide range of land monitoring applications \cite{miller2024deep}. However, exploiting multi-modal SITS also presents unique challenges due to satellite acquisition constraints. Time series frequently exhibit irregular temporal sampling due to cloud coverage, and varying acquisition scenarios.
Moreover, due to orbit patterns, observations are often temporally unaligned, as SITS may have various acquisition dates. 
Despite these challenges, many time-critical applications, such as deforestation, crop health or economic activity monitoring require regular and timely predictions. Ideally, each new satellite acquisition should immediately trigger updated predictions for the observed area, enabling rapid decision-making for environmental management and policy implementation. Thus, in this work we study auto-regressive methods that can handle SITS specificities, perform effective multi-sensor data fusion, capture long-range temporal dependencies, enable parallel sequence processing during training, and maintain computational efficiency for real-time inference.

Following the success of fully attentional architectures such as the Transformer \cite{Vaswani2017} in Natural Language Processing (NLP), these architectures have been widely adapted for SITS analysis \cite{russwurm20_self_atten_raw_optic_satel,garnot2020satellite,dumeur-2024-self-super}. Based on scaled dot-product attention, these architectures enable efficient parallel training compared to earlier recurrent neural network approaches. Moreover, through their positional encoding mechanisms, they can handle SITS temporal irregularity and unalignment. Furthermore, Transformers have demonstrated superior performance in handling long temporal dependencies and fusing multi-modal data \cite{chameleonteam2025chameleonmixedmodalearlyfusionfoundation}.
Nevertheless, for regular and up-to-date land monitoring tasks, a major limitation remains: analyzing a novel satellite image with Transformers requires reprocessing the entire time series, resulting in computationally expensive inference that prevents the regular monitoring over large areas. This constraint is increased by the Transformer's quadratic computational complexity with respect to sequence length.

Furthermore, in Natural Language Processing, linear/kernel attention mechanisms \cite{pmlr-v119-katharopoulos20a,qin2022cosformer} as well as retention mechanisms \cite{sun2023retentive} have emerged as alternatives to Transformer attention. These mechanisms offer two complementary advantages: they preserve the parallelizable training characteristics of Transformers while enabling recurrent inference, where new inputs can be processed incrementally without recomputing the entire sequence. As in~\cite{sun2023retentive} we refer to these methods as \textit{dual-form}, as they allow parallel training and are recurrent at inference. In practice, this is achieved by replacing the softmax in transformer attention.
Specifically, the first kernel-based attention network, the Linear Transformer \cite{pmlr-v119-katharopoulos20a}, simply replaced softmax with a separable similarity function. Subsequently, with the CosFormer, \cite{qin2022cosformer} demonstrated that performances in NLP tasks could be improved by integrating cosine-based reweighting based on token distances. In the retention mechanism proposed in Retentive Neural Networks~\cite{sun2023retentive}, the reweighting mechanism is inspired by Rotary Positional Encoding (RoPE) \cite{su2024roformer}.
These architectures exclusively support the causal form of attention (relying only on past and current tokens), which is the only formulation that allows recurrent processing where future tokens are unavailable.
Despite their potential, these mechanisms have mostly been applied to NLP tasks and have not yet been explored for SITS analysis, let alone for multi-modal time series.

Consequently, this study aims to fill this knowledge gap by comparing various kernel-based attention mechanisms: the Linear Transformer \cite{pmlr-v119-katharopoulos20a}, CosFormer \cite{qin2022cosformer}, the RoPE Linear Transformer~\cite{su2024roformer}, and Retention \cite{sun2023retentive}. Due to SITS specificities, we also propose modifications of these existing architectures to address SITS temporal irregularity unalignment. 
We evaluate these dual form mechanisms using Sentinel-1 (S1) and Sentinel-2 (S2) SITS on two supervised tasks. First, due to the scarcity of labels for multi-temporal and multi-modal remote sensing tasks, we conduct experiments on a forecasting task that serves as a proxy task requiring no manual labels. Second, we evaluate the various dual form mechanisms on a real-world application: solar panel construction site monitoring. 
In summary, the contributions of this paper are as follows:

\begin{itemize}
    \setlength\itemsep{0em}
    \item A unified formulation of dual form mechanisms for SITS analysis;
    \item A novel multi-modal spectro-spatio-temporal architecture that enables parallel training and efficient inference for land monitoring applications;
    \item And comprehensive experimental results demonstrating that dual form mechanisms are effective for multi-modal SITS analysis across both forecasting and multi-temporal segmentation tasks.
\end{itemize}

\section{Dual-form neural networks}\label{sec:dual-archi}
In this paper, we explore both linear attention and retention mechanisms for efficient auto-regressive multi-modal SITS analysis. Both mechanisms refer to a type of sequence analysis, where, given an input sequence $X=[\mathbf{x}_1,\ldots \mathbf{x}_T]$, composed of T vectors $\mathbf{x}_i \in \mathbb{R}^{(1,d)}$, each output vector $\mathbf{o}_i$ corresponds to a weighted sum of the value vectors $(\mathbf{v}_j=\mathbf{x}_j W_V \in \mathbb{R}^{(1,d_V)})_{1\leq i \leq T}$,
\begin{equation}\label{eq:attn-ret-general}
    \mathbf{o}_i=\sum_{j=1}^i a_{i,j} \mathbf{v}_j.
\end{equation}
While retention and linear attention differ in the similarity score $a_{i,j}$ used, they both exhibit parallel and recurrent formulations that are suitable for fast training and efficient inference, respectively. This section first reviews the similarity scores $a_{i,j}$ for both attention and retention mechanisms before detailing their dual formulations.

\subsection{Similarity scores}
\subsubsection{Linear attention}\label{sec:similarity-fn}
In causal attention, the similarity scores $a_{i,j}$ correspond to $\frac{\text{s}(\mathbf{q}_i,\mathbf{k}_j)}{\sum_{j=1}^i \text{s}(\mathbf{q}_i,\mathbf{k}_j)}$, where $\text{s}$ is the similarity function and $\mathbf{q}_i=\mathbf{x}_i W_Q \in \mathbb{R}^{(1,d_K)}$ and $\mathbf{k}_j=\mathbf{x}_j W_K \in \mathbb{R}^{(1,d_K)}$ correspond to the query and key vectors, respectively, thus the output vectors write as
\begin{equation}\label{eq:general-attn}
\mathbf{o}_i=\frac{\sum_{j=1}^{i}\text{s}\left(\mathbf{q}_i,\mathbf{k}_j\right) \mathbf{v}_j}{\sum_{j=1}^{i} \text{s}\left( \mathbf{q}_i, \mathbf{k}_j\right)}.
\end{equation}
A crucial component of attention is the similarity function $\text{s}(\cdot,\cdot)$ employed. For instance, the Transformer architecture \cite{Vaswani2017} uses the exponential of the scaled dot product to compute similarity between pairs of query and key vectors
\begin{equation}\label{eq:vaswani-attn}
    \text{s}(\mathbf{q}_i,\mathbf{k}_j)=\exp\left(\frac{\mathbf{q}_i \cdot \mathbf{k}_j^T}{\sqrt{d_K}}\right).
\end{equation}
In contrast, linear attention mechanisms rewrite attention through the lens of kernel methods \cite{tsai-etal-2019-transformer}. The core idea introduced by \cite{pmlr-v119-katharopoulos20a} is to employ a separable similarity kernel \begin{equation}\label{eq:separable-sim}
    \text{s}(\mathbf{q}_i,\mathbf{k}_j)=\phi(\mathbf{q}_i)\cdot\phi(\mathbf{k}_j)^T,
\end{equation} where $\phi$ corresponds to the feature map function, thus allowing both parallel and recurrent forms of causal attention (see \cref{sec:dual-form}). As shown in \cref{tab:sim-fns}, existing linear attention mechanisms differ in the $\phi$ function employed. A positive function $\psi$ is used to map the key and query vector to $\mathbb{R}^{+^{(1,d_K)}}$.

\begin{table*}[htbp]
    \centering
    \begin{tabular}{lc}
        \toprule
        Attention mechanism  & Similarity function $\text{s}(\mathbf{q}_i,\mathbf{k}_j)$\\
        \midrule
        \addlinespace[5pt]
        Transformer \cite{Vaswani2017}& $\exp({\nicefrac{\mathbf{q}_i \cdot \mathbf{k}_j}{\sqrt{d_K}}})$ \\
        \addlinespace[3pt]
        Linear Transformer \cite{pmlr-v119-katharopoulos20a} & $\left(1+\text{elu}(\mathbf{q}_i)\right)\cdot \left(1+\text{elu}(\mathbf{k}_j)\right)^T$\\
        \addlinespace[3pt]
        CosFormer \cite{qin2022cosformer} &  $\cos(\frac{\pi}{2} \times\frac{i-j}{M})\psi(\mathbf{q}_i)\cdot\psi(\mathbf{k}_j)^T$\\
        \addlinespace[3pt]
        RoPE Linear transformer \cite{su2024roformer} & $(\psi(\mathbf{q}_n) R_{i\boldsymbol{\theta}})(\psi(\mathbf{k}_m) R_{j\boldsymbol{\theta}})^T$ \\ 
        \addlinespace[2pt]
        \bottomrule
    \end{tabular}
    \caption{Considered attention mechanisms in this study and corresponding similarity functions.}
    \label{tab:sim-fns}
\end{table*}

First, the linear Transformer~\cite{pmlr-v119-katharopoulos20a} achieves impressive results in NLP by simply using $\phi^{\text{linear}}(\mathbf{u})=1+\text{elu}(\mathbf{u})$. More recent architectures, such as CosFormer~\cite{qin2022cosformer}, propose integrating a reweighting mechanism that strengthens the influence of nearby tokens in the attention while decreasing the importance of distant elements. More specifically, CosFormer employs cosine-based reweighting with $\cos\left( \frac{\pi}{2} \times \frac{i-j}{M}\right)$, where $M$ is the maximum sequence length and the cosine values range between 0 and 1. Precisely, to match the general notation of linear attention:
\begin{align}
    \phi^{\text{cosformer}}(\mathbf{u}_i)=&\;[\cos(\frac{\pi i }{2 M})\psi(\mathbf{u}_i),\sin(\frac{\pi i }{2 M})\psi(\mathbf{u}_i)] \\
    \text{s}^{\text{cosformer}}(\mathbf{q}_i,\mathbf{k}_j)=&\;\phi^{\text{cosformer}}(\mathbf{q}_i)\phi^{\text{cosformer}}(\mathbf{k}_j)^T \\
    \text{s}^{\text{cosformer}}(\mathbf{q}_i,\mathbf{k}_j)=&\;\cos(\frac{\pi i }{2 M})\psi(\mathbf{q}_i)\cos(\frac{\pi j }{2 M})\psi(\mathbf{k}_j)^T +  \\&\;\sin(\frac{\pi i }{2 M})\psi(\mathbf{q}_i)\sin(\frac{\pi j }{2 M})\psi(\mathbf{k}_j)^T.
\end{align}
While this separable reweighting mechanism  achieves great results in NLP, it exhibits potential drawbacks. First, it requires knowledge of a maximum allowed sequence length M during inference. Second, it cannot model periodic events, as the weights decrease with the distance between tokens.
An alternative to cosine-based reweighting is the use of Rotary Positional Encoding (RoPE)~\cite{su2024roformer}. In this mechanism, the query $\mathbf{q}_i$ and key vector $\mathbf{k}_j$ are projected by rotary matrices $R_{\boldsymbol{\theta},i}$ and $R_{\boldsymbol{\theta},j}$, respectively. Thus, we consider $\phi^{\text{RoPE}}(\mathbf{u}_i)=\psi(\mathbf{u}_i)R_{\boldsymbol{\theta},i}$. As detailed in \cref{ap:rotary}, pairs of features for query and key are rotated by an angle $i \times \theta_k $ that is proportional to the position $i$ of the token in the sequence. Since $R_{\boldsymbol{\theta},i} R_{\boldsymbol{\theta},j}^T=R_{\boldsymbol{\theta},(i-j)}$, RoPE enables the integration of relative position encoding within the attention mechanism. While RoPE does not rely on a sequence length $M$, it may produce negative attention scores within the sequence. In the remainder of this manuscript we refer to the application of RoPE with a linear attention mechanism as LinRoFormer.

\subsubsection{Retention scores}
At the core of the Retentive neural network is the retention mechanism, which shares similarities with LinRoFormer by integrating relative distances between tokens as rotations of the features
\begin{equation}\label{eq:retention}
    a_{i,j} = \gamma_h^{i-j}(\psi(\mathbf{q}_i) R_{\boldsymbol{\theta},i})(\psi(\mathbf{k})_jR_{\boldsymbol{\theta},j})^T.
\end{equation}
Nevertheless, two important differences should be highlighted. First, a decay term $\gamma_h \in [0,1]$ is integrated, which varies for each retention head (the counterpart of attention heads) and is expected to stabilize the RoPE integration (similar to xPOS encoding \cite{sun-etal-2023-length}). Second, in contrast to attention scores where $\sum_m a_{n,m}=1$, no normalization of the scores is performed in retention.

\subsection{Dual formulations}\label{sec:dual-form}
All previously described mechanisms have the ability to process a sequence $X$ into a sequence $O=[\mathbf{o}_1,\ldots,\mathbf{o}_N]$ in either:
\begin{itemize}
    \setlength\itemsep{0em}
    \item \textbf{Parallel form:} $O=\sigma(M)V$, where all tokens ($\mathbf{v}_i=\mathbf{x}_i W_V$) are processed in parallel to output all target tokens $\mathbf{o}_i$ and $\sigma$ may be a scaling operation or the identity function. 
    \item \textbf{Recurrent form:} $\mathbf{o}_n= \text{r}(\mathbf{h}_{n-1},\mathbf{x}_n)$, where the output $\mathbf{o}_n$ is computed using a hidden state $\mathbf{h}_{n-1}$ and the current token $\mathbf{x}_n$.
\end{itemize}
Both linear attention and retention implement a multi-head mechanism, where specifically query, key and value vectors are split along the channel dimension to produce $n_{head}$ triplets $(\mathbf{q}_h,\mathbf{k}_h ,\mathbf{v}_h)$ with $\mathbf{q}_h,\mathbf{k}_h \in \mathbb{R}^{(1,d_K/n_{\text{head}})}$ and $\mathbf{v}_h\in \mathbb{R}^{(1,d_V/n_{\text{head}})}$. While the final output of the linear attention mechanisms is $\mathbf{o}=[\mathbf{o}_1; \ldots \mathbf{o}_h\ldots ;\mathbf{o}_{n_{\text{heads}}}] \in \mathbb{R}^{(1,d_v)}$, the multi-head retention mechanism, normalizes the output of each heads with a group Norm 
\begin{align}\label{eq:multiscaleret}
    \mathbf{o}^{norm}&=\text{GroupNorm}_h\left([\mathbf{o}_1; \ldots \mathbf{o}_h\ldots ;\mathbf{o}_{n_{\text{heads}}}]\right) \\
    \mathbf{o}&=\text{swish}(\mathbf{x}W_G \odot \mathbf{o}^{norm})W_O,
\end{align} with $W_G, W_O$ learnable matrices both in $\mathbb{R}^{(d_{\text{model}},d_{\text{model}})}$.
To simplify notation, parallel and recurrent forms are detailed on one single head.

\paragraph{Parallel formulation.}
First, all mechanisms can be written under the same parallel formulation \begin{equation}\label{eq:parallel}
    M=L\odot \phi (Q) \phi(K)^T,
\end{equation} where $L$ corresponds to a lower triangular matrix and $\odot$ the Hadamard product.
While $L$ corresponds to a causal masking matrix with $l_{i,j}=1$ for linear attention, it encodes decay information in retention with $l_{i,j}=\gamma^{i-j}$. In addition, in contrast to the retention mechanism, the attention mechanism performs a row-wise normalization operation noted as $\sigma(\cdot)$. This formulation allows fast training, as all outputs can be computed in a single pass.

\paragraph{Recurrent formulation.}
The recurrent forms of attention and retention also differ slightly due to the normalization process in attention.
Therefore, linear attention requires to store two hidden states $T^{\text{attn}}_n$ for the accumulation, and $\mathbf{d}_n$ for the normalization:
\begin{itemize}
    \setlength\itemsep{0em}
    \item[ ] $T^{\text{attn}}_n \in \mathbb{R}^{(d_K,d_v)}$ with $T^{\text{attn}}_{n+1}=T^{\text{att}}_n+\phi(\mathbf{k}_{n+1})^T \mathbf{v}_{n+1}$
    \item[ ] $\mathbf{d}_n \in \mathbb{R}^{d_K}$ with $\mathbf{d}_{n+1}=\mathbf{d}_n+\phi(\mathbf{k}_{n+1})$.
\end{itemize}
Using these two hidden states, the output of the attention for the image $n+1$ can be written as
\begin{align} \label{eq:linear-attn-rec}
\begin{split}
\mathbf{o}^{\text{attn}}_{n+1}=\phi(\mathbf{q}_{n+1})\left[T^{\text{attn}}_n + \phi(\mathbf{k}_{n+1})^T \mathbf{v}_{n+1}\right] \\ \oslash \left(\phi(\mathbf{q}_{n+1})\left[\mathbf{d}_n + \phi(\mathbf{k}_{n+1})\right]\right),
\end{split}
\end{align}
with $\oslash$ corresponding to scalar division and $\left(\phi(\mathbf{q}_{n+1})\left[\mathbf{d}_n + \phi(\mathbf{k}_{n+1})\right]\right)$ a scalar corresponding to the attention normalization. 

In contrast, retention only stores the hidden state $T_n$, which is multiplied by a decay factor $\gamma \in [0,1]$ at each time step
\begin{equation}
    T^{\text{ret}}_{n+1}=\gamma T^{\text{ret}}_n+\phi(\mathbf{k}_{n+1})^T \mathbf{v}_{n+1}
\end{equation}
and
\begin{equation}
    \mathbf{o}^{\text{ret}}_{n+1}=\phi(\mathbf{q}_{n+1})  [\gamma T^{\text{ret}}_n + \phi(\mathbf{k}_{n+1})^T\mathbf{v}_{n+1} ].
\end{equation}
These dual formulations appear promising for SITS analysis, offering both embarrassingly parallel learning on historical sequences and recurrent inference with constant memory for processing new acquisitions. However, these architectures were initially developed for natural language processing (NLP) tasks. Although they offer performance comparable to that of causal transformers in NLP applications, their adaptation to multi-modal satellite image time series remains an open question, which this study examines.
\section{Method}\label{sec:method}
This section first details the proposed multi-modal SITS encoder that encapsulates the various \df mechanisms, then describes the two supervised tasks investigated in this study. Finally, the datasets used for the experiments are defined.

\subsection{Architecture} \label{sec:archi}
\subsubsection{Backbone overview}

\begin{figure}
\centering
\includesvg[width=\linewidth]{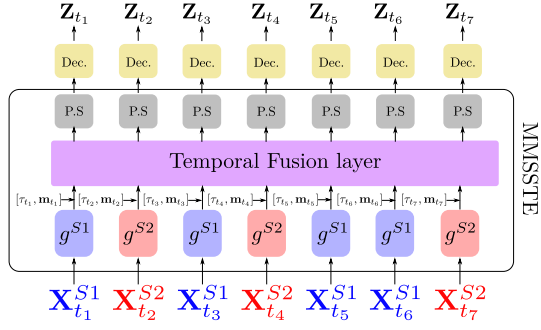}
\caption{Overview of the proposed multi-modal spectro-spatio temporal architecture, composed of: a modality specific spectro-spatial encoder (noted $g^{S1}$ resp. $g^{S2}$), a temporal fusion encoder, an upsampling operation with pixel shuffle (P.S) and a task specific decoder layer.}  
\label{fig:archi-overview}
\end{figure}

The core of this paper is the comparison of various auto-regressive mechanisms for multi-modal SITS analysis.
To establish the comparison of various auto-regressive mechanisms on tasks that require live predictions, we design a novel architecture focused on the encoding of multi-modal data with temporal information. The proposed multi-modal spectro-spatio-temporal encoder (MMSSTE) is inspired by \cite{dumeur-2024-self-super,garnot-2021-panop-segmen}, and is illustrated in \cref{fig:archi-overview}. It is composed of three main blocks:
\vspace*{-0.5em}
\begin{itemize}
    \setlength\itemsep{0em}
\item A modality-specific spectral-spatial downsampling encoder (SSE), which processes each image of the SITS independently. Given an image $\mathbf{X}_t \in \mathbb{R}^{(C,H,W)}$, the SSE outputs a feature map $\mathbf{X}_{(t,\cdot, \cdot,\cdot)}' \in \mathbb{R}^{(d_{\text{model}},H/2,W/2)}$. We denote $g^{S1}$ and $g^{S2}$ the encoders for Sentinel-1 and Sentinel-2 respectively.
\item A multi-modal temporal fusion layer which processes independently the $(H/2) \times (W/2)$ sequences $\mathbf{X'}_{(\cdot,\cdot,h,w)} \in \mathbb{R}^{(T,d_{\text{model}})}$ using one of the multi-head attention mechanisms.
\item A spatial up-sampling (P.S) layer which up-samples each of the $T$ processed acquisitions independently to $ \mathbf{Y}_{(t,\cdot,\cdot,\cdot)} \in \mathbb{R}^{(d_{\text{model}},H,W)}$ using a pixel shuffle operator~\cite{shi2016real}.
\end{itemize}\vspace*{-0.3em}

A task-specific decoder that projects $\mathbf{Y}_t$ into the $T$ predicted targets $\mathbf{Z}_t \in \mathbb{R}^{(K,H,W)}$ is used in the two considered supervised training tasks.
In this study, various \df mechanisms are utilized to capture the synergies between modalities as well as extract complex temporal patterns. Other parts of the architecture (the SSE, spatial up-sampling, and task-specific decoder) remain the same. More specifically, the SSE employed is a U-Net with four down-sampling blocks and only three up-sampling layers, which results in a spatial down-sampling operation.
Then, positional encoded acquisition dates information $(\text{PE}(\tau_{t_i}))_{i\leq T}$\footnote{We employ the absolute positional encoding as proposed in the original Transformer \cite{Vaswani2017}.} and their corresponding modality specific tokens $(\mathbf{m}_{t_i})_{i\leq T}$ are concatenated to their corresponding sequence token $(\mathbf{X'}_{(t_i,\cdot,h,w)})_{i\leq T}$, before being processed by the multi-modal multi temporal encoder. Precisely, the multi-modal fusion layers is composed of $n_{\text{layers}}$ layers. Each layer is composed of an attention (resp. retention) mechanism followed by feed-forward layers.
The following hyper-parameters are used: $n_{\text{layers}}=3$, $n_{\text{heads}}=4$, $d_K=64$, and $d_{\text{model}}=64$.
After the attention mechanism, the spatial up-sampling is performed with a pixel shuffle operation~\cite{shi2016real}.

\subsubsection{Temporal fusion layers}
Existing \df mechanisms, previously described in \cref{sec:dual-archi}, were originally developed for NLP and, to the best of our knowledge, have never been applied to SITS. Consequently, they mostly rely on token indices in the sequence and do not address SITS temporal irregularity and unalignment.  Therefore, in this paper we propose variations of the \df mechanisms that compute distances between tokens based on actual temporal information. Specifically, modified versions of CosFormer, LinRoFormer, and the Retention neural network are proposed and are respectively named Time CosFormer, Time LinRoFormer, and Time Retention.

In these proposed architectures, for a pair of tokens $\mathbf{x}_{i}, \mathbf{x}_j$, instead of computing the distance between token indices $i-j$, we employ $t_i-t_j$, which represents the distance in days between acquisitions. For Time CosFormer, we employ $M=700$, meaning that we consider that the maximum considered temporal distance between any two acquisitions in our tasks is 700 days. 
As the feature map function, we always employ $\psi(\mathbf{x})=\text{elu}(\mathbf{x})+1$, even if the original CosFormer~\cite{qin2022cosformer} uses $\psi(\mathbf{x})=\text{ReLU} (\mathbf{x})$ and original retention~\cite{sun2023retentive} $\psi(\mathbf{x})=\mathbf{x}$, as it appears to avoid head collapsing and stabilizing training.
Finally, the training stability of LinRoFormer (resp. Time LinRoFormer) and Retention (resp. Time Retention) was improved by scaling $\phi^{\text{RoPE}}(\mathbf{u})/d_K$, where $d_K$ is the number of features per head.

\subsection{Supervised training tasks}
The investigated kernel-based linear attention mechanisms are compared on both a SITS forecasting task and a multi-temporal segmentation task. The forecasting task corresponds to a proxy task, which allows assessment of the model's ability to extract complex temporal understanding of the SITS. In contrast, the multi-temporal segmentation task corresponds to a realistic remote sensing application.
In this section, we describe how the outputs of the MMSSTE backbone are used depending on the task.

\subsubsection{Multi-modal forecasting task}
\begin{figure}
\centering
\includesvg[width=\linewidth]{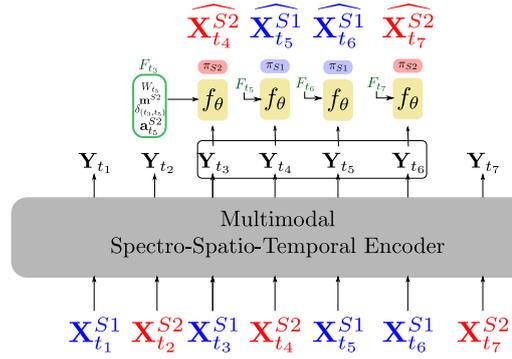}
\caption{Multi-modal forecasting framework.}
\label{fig:forecast-overview}
\end{figure}

In the forecasting task, given the latent representation at time-step $t$ denoted $\mathbf{Y}_t$, the decoder predicts $\widehat{\mathbf{X}}_{t+1}$. As shown in \cref{fig:forecast-overview}, a forecaster model $f_{\theta}$ is placed on top of the output of the MMSSTE and processes independently the pixel-level representation of the multi-modal time series at step $t$ denoted $\mathbf{Y}_{(t,\cdot,h,w)} \in \mathbb{R}^{d_{\text{model}}}$ (where $h,w$ is the location of the pixel). In other words, the forecaster only processes the channel dimension and does not capture temporal nor spatial relationships.

The same forecaster is used to reconstruct either S1 or S2 data. Modality-specific linear heads denoted $\pi^{S1}$ (resp. $\pi^{S2}$) are employed on top of the forecaster to enforce the correct number of channels of the reconstruction. Along with the latent representation of the pixel-level time series at time-step $t$ to predict $\mathbf{X}_{(t+1,\cdot ,h,w)}$, auxiliary variables are employed. More specifically, we integrate:
\begin{itemize}
    \setlength\itemsep{0em}
    \item Weather AgERA5~\cite{agera5} time series of the $n_w$ days preceding date $t+1$ denoted $W_{t+1} \in \mathbb{R}^{(n_w,d_w)}$, where $d_w$ is the number of AgERA5 variables.
    \item A learnable modality-specific token named $\mathbf{m}^{S1}$ (resp. $\mathbf{m}^{S2}$).
    \item Angle variables $\mathbf{a}^{S1}$ for S1 data containing an encoding of azimuth and incidence angles. For S2 forecasting, $\mathbf{a}^{S2}$ is just a learnable placeholder.
    \item The distance in days between the previous acquisition $t$ and the one to predict denoted $\delta_{(t,t+1)}$.
\end{itemize}

The encoding of the auxiliary variables $W_{t+1}$ and $\mathbf{a}^{S1}_t$ is detailed in \cref{app:aux-encoding}.
To incorporate relative distance between predictions, we integrate rotary positional encoding (RoPE) \cite{su2024roformer}. The RoPE performs rotations of the features of the latent representation $\mathbf{Y}_{(t,\cdot,h,w)}$ by an angle proportional to the temporal distance $\delta_{(t,t+1)}$. It relies on a rotary matrix $R_{\boldsymbol{\theta},\delta_{t,t+1}} \in \mathbb{R}^{(d_{\text{model}},d_{\text{model}})}$
\begin{equation}\label{eq:rope}
    \text{RoPE}(\mathbf{Y}_{(t,\cdot,h,w)},\delta_{t,t+1})=R_{\boldsymbol{\theta},\delta_{t,t+1}} \cdot \mathbf{Y}_{(t,\cdot,h,w)}.
\end{equation}
Then, once the rotary positional encoding is applied, the resulting encoded representation is concatenated with auxiliary variables and processed by an MLP as
\begin{equation}\label{eq:decoder}
    \mathbf{Z}_{t+1}^{S1}=\text{MLP}\left([\text{RoPE}(\mathbf{Y}_t,\delta_{t,t+1});\mathbf{m}^{S1};f_w(W_{t+1});f_{\theta}(\mathbf{a}_{t+1})]\right).
\end{equation}
The training loss corresponds to a weighted sum of the Mean Square Error (MSE) losses on each S1 and S2 data
\begin{equation}
    L=\text{MSE}_{S1}+w_{S2}\text{MSE}_{S2}.
\end{equation}
The loss is only applied on time steps $t \in [n_\text{after}, T]$. In all our experiments, we set $n_\text{after} = 6$ and $w_{S2}=0.1$.
Validity and cloud masks are provided to the reconstruction loss to allow the model to learn to ignore clouds and out-of-swath regions.
\subsubsection{Multi-temporal segmentation task}
In the multi-temporal segmentation task, multiple labels are provided for each time series. At some time steps, labels $Z_t$ are provided. To perform this real-world application scenario, a linear classification layer is placed on top of the MMSSTE to project the output of each representation into the correct number of classes. The model is trained using Focal Loss. While the model outputs predictions for all time steps, the loss is computed only on time steps with available labels.

\subsection{Datasets}\label{sec:dataset}

To support the comparison of attention mechanisms, we collect one dataset for each task.
Both datasets comprise of Sentinel-1 (SAR) and Sentinel-2 (optical) time series.
For S1, IW GRD DV products are selected and processed with thermal noise removal and gamma radiometric calibration, and geocoded to 10m resolution using the GLO-30 digital elevation model. Consecutive slices are assembled, and the amplitudes are saved with linear encoding.
The validity masks are defined as the in-swath areas.
For S2, bottom-of-atmosphere L2A products are used, selecting the 10m and 20m (upsampled to 10m) bands. Products with a cloud cover larger than 20\% are excluded from the dataset.
Besides the in-swath areas, the validity masks for S2 also excludes regions classified as cloud or shadow by the CloudSEN12 cloud detector~\cite{aybar2022cloudsen12}.

\paragraph{Multi-modal SITS forecasting dataset.}

The forecasting dataset is composed of multi-modal SITS and exhibit significant temporal and geographical diversity. Patches of size $512 \times 512$ pixels at 10m resolution have been selected throughout the world in various eco-regions and urban and sub-urban areas. In total, 640 sites were sampled. For each site, 15 consecutive months of S1 and S2 data is downloaded, within the range 2019--2021. The resulting training set contains 444 samples, and 96 additional sites for the validation and for the test set each.

For the weather data, AgERA5~\cite{agera5} time series of $n_w=10$ past days are extracted for each satellite acquisition, with the following variables ($d_w=8$): precipitation flux, solar radiation flux, vapour pressure, 2m dewpoint temperature, and 10m wind speed; each using the 24h mean statistic, and 2m temperature with min/max/mean statistic.

\paragraph{Multi-modal solar panel segmentation dataset.}

This supervised segmentation dataset relies on human annotations. 650 solar panel projects were labeled over their duration using S2 imagery, totaling about 13000 annotated masks. Masks are composed of three regions: un-constructed, installed solar panel, and out-of-site regions.
To complement the S2 acquisitions, S1 is downloaded but not labeled.
The resulting multi-modal dataset is split into 8 folds to perform k-fold training. For a given experiment the training is composed of five folds, one in validation and tow on tests.

\section{Experiments and Results} \label{sec:expe-res}
The proposed model takes SITS of spatial dimension $(128 \times 128)$ as input. Both multi-modal tasks are compared with their S2 mono-modal versions. For training efficiency, S1 and S2 SITS lengths are limited to 16 acquisitions each during training and validation.
For the forecasting task, models are trained for 1400 epochs on a single NVIDIA A100 GPU with a batch size of 4 and a learning rate of 0.001. Four training runs are performed on different training set splits, resulting in testing on different test sets.
For the solar panel detection task, models are trained for 600 epochs on a single NVIDIA V100 GPU with a batch size of 2, a learning rate of 0.006, and focal loss parameters ($\beta=0.58, \gamma=2.0$). Using the solar panel datasets k-fold split, two different experiments with different training, validation and test split are performed.
Mean and standard deviation of the results are reported for both tasks. We first present results on the forecasting task, which serves as a proxy task for complex temporal feature extraction, before detailing results obtained on a real-world application (solar panel detection).
\subsection{Multi-modal forecasting}\label{res:sits-forecast}
\begin{figure*}[!ht]
    \centering
    \includegraphics[width=\linewidth]{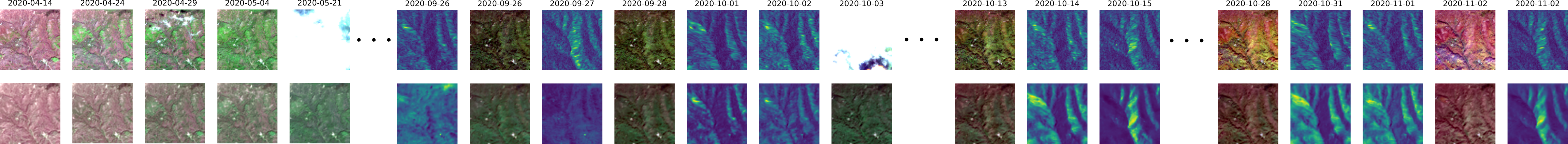}
    \caption{Visualization of the reconstruction in the multi-modal forecasting task using the Time CosFormer mechanism. The top row corresponds to the input and target SITS. The $(n-1)^{\text{th}}$ images are used by the model to predict the $n^{\text{th}}$ one. The bottom row corresponds to the reconstruction by our model.}
    \label{fig:forecast-visu}
\end{figure*}

\begin{figure}[!htb]
    \centering
    \includegraphics[width=\linewidth]{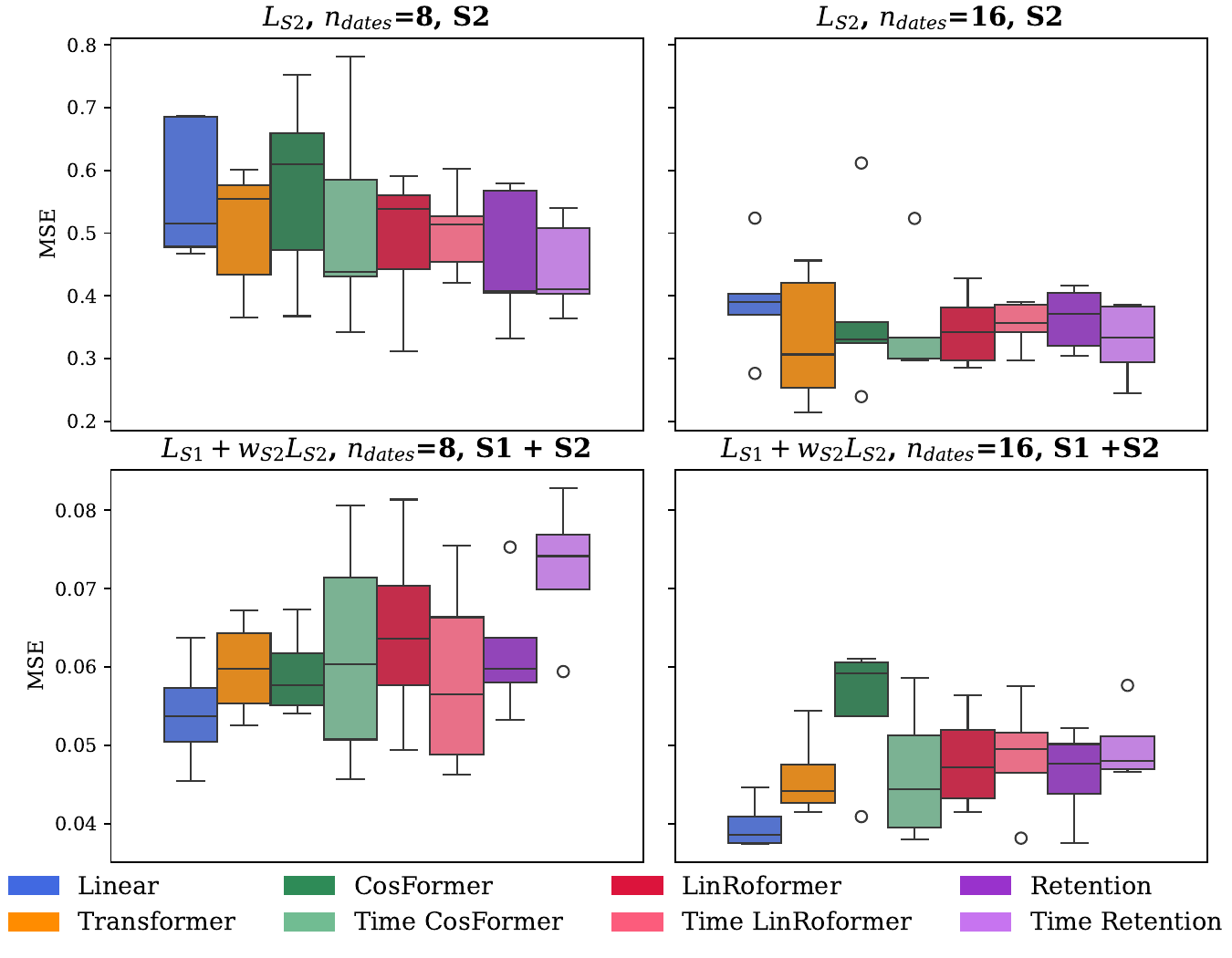}
    \caption{Boxplot showing the MSE on the forecasting task for various temporal fusion encoders. Top row represents result obtained in the mono-modal S2 forecasting framework, bottom row corresponds to the multi-modal reconstruction loss. On the boxplot a line is drawn at the median value.}
    \label{fig:boxplot-forecast}
\end{figure}

We first assess the various \df mechanisms on a forecasting task with $n_{\text{dates}}=16$ for each modality (consistent with the training and validation framework) and with $n_{\text{dates}}=8$. In this first experiment, images are not selected consecutively; thus, time series may exhibit significant temporal gaps within the input sequences. The reconstruction errors for both mono-modal and multi-modal frameworks are shown in \cref{fig:boxplot-forecast}.

In the framework consistent with training conditions ($n_{\text{dates}}=16$), for both multi-modal and mono-modal frameworks, the average MSE values are close. When the number of dates within the SITS is small, we observe stronger disparities in the results compared to the training framework with $n_{\text{dates}}=16$. As expected, shorter input SITS (Satellite Image Time Series) result in more challenging predictions. Within the mono-modal framework ($n_{\text{dates}}=8$), dual mechanisms that integrate temporal information, specifically the Time CosFormer, Time Retention, and Time LinRoFormer, consistently outperform their original counterparts. Furthermore, in mono-modal forecasting, linear mechanisms incorporating a reweighting scheme often slightly outperform both the original Transformer and standard linear attention.
\begin{figure*}[!hbt]
    \centering
    \includegraphics[width=\linewidth]{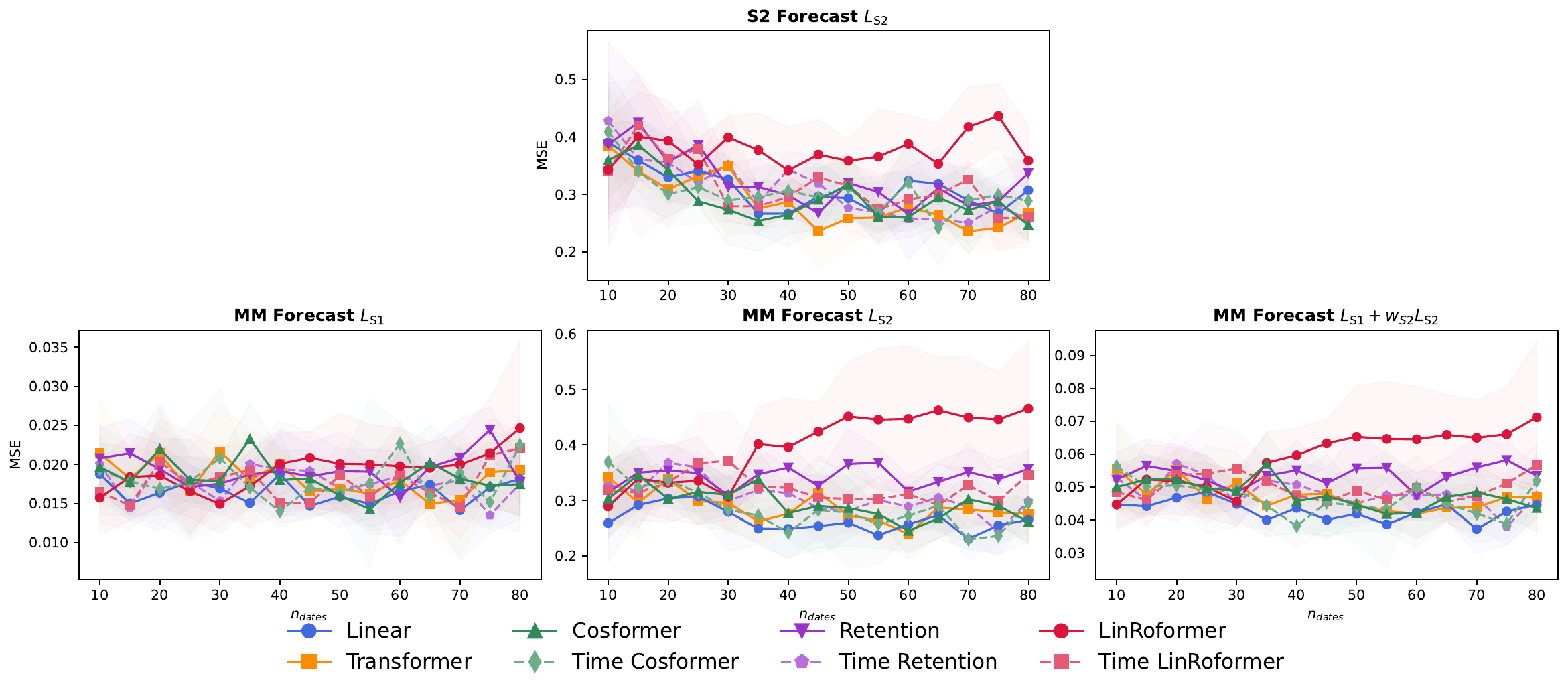}
    \caption{Effect of the look-back length in the forecasting task for various \df mechanisms. Mono-modal results are displayed on the top-line and multi-modal results on the bottom line.}
    \label{fig:look-back-forecast}
\end{figure*}

Surprisingly, in the multi-modal configuration, linear attention outperforms all other mechanisms. We hypothesize that because the reweighting mechanism is based solely on the relative temporal distance between acquisitions, it fails to account for the fact that tokens within the sequence belong to different modalities. In other words, when processing a token at timestep $t_i$, the most relevant information may not reside in the previous acquisition, but rather in the closest acquisition of the same modality.

Lastly, importantly, for both mono-modal and multi-modal tasks, the dual mechanisms appear to have similar performance to Transformer attention while exhibiting a recurrent form. 

We have also confirmed the quality of the forecasting task by qualitatively looking at the reconstructions. As shown in \cref{fig:forecast-visu}, we observe that the model is able to extract and summarize sufficient past information in the representation $\mathbf{Y}_t$ to allow the forecaster to forecast $\mathbf{Y}_{t+1}$.

\subsection{Look-back length effect}

We study in \cref{fig:look-back-forecast} the performance of various \df mechanisms on either mono-modal (S2) or multi-modal forecasting tasks as a function of the look-back length. 
Specifically, for each SITS in the test dataset, we select consecutive $n_{\text{dates}}$ acquisitions and ensure that experiments with higher $n_{\text{dates}}$ values include images from experiments with smaller $n_{\text{dates}}$ values. If $n_{\text{dates}}$ is larger than the maximum length of S1 or S2, we set $n_{\text{dates}}=\min(n_{S1},n_{S2})$, corresponding to the minimum between the S1 and S2 lengths.
First, for most models, the S2 MSE in either mono-modal or multi-modal forecasting decreases with an increase in $n_{\text{dates}}$. The noise contained in S1 data prevents us from observing such a clear trend in $L_{S1}$ in \cref{fig:look-back-forecast}.
Then, in the multi-modal task, apart from the LinRoFormer and Retention, we observe that most \df mechanisms exhibit an improvement in $L_{S2}$ with an increase in look-back length. This illustrates interesting generalization capabilities, while these mechanisms have been trained with SITS of a maximum length of 16, they can be used at inference on much longer sequences. Although LinRoFormer and Retention achieve good results under settings similar to training conditions (see \cref{res:sits-forecast}), they have the highest S2 reconstruction errors in the multi-modal setting. This suggests that reweighting in dual-form mechanisms based on RoPE may generalize better if they rely on temporal acquisition dates.
Second, we observe that other mechanisms (Time Retention, Time LinRoFormer, Linear Transformer, CosFormer and Time CosFormer) achieve results comparable to the transformer with causal attention (orange curve). Consequently, for the forecasting task, the similarity function in a causal Transformer can be replaced with other \df mechanisms (kernel-based attention or retention) without performance degradation.
In terms of \df mechanism comparison, for those that integrate a cosine re-weigthing mechanism (CosFormer and  Time CosFormer), we observe that they appear to slightly improve the multi-modal reconstruction results compared to Time Retention and Time LinRoFormer. Lastly, consistent with our previous experiments, we observe that linear attention outperforms other dual mechanisms in the multi-modal forecasting task. This highlights the limitations of existing dual mechanisms that incorporate temporal re-weighting when applied to multi-modal data.

\subsection{Multi-temporal solar panel detection}
\begin{figure*}[htb!]
    \centering
    \includegraphics[width=0.95\linewidth]{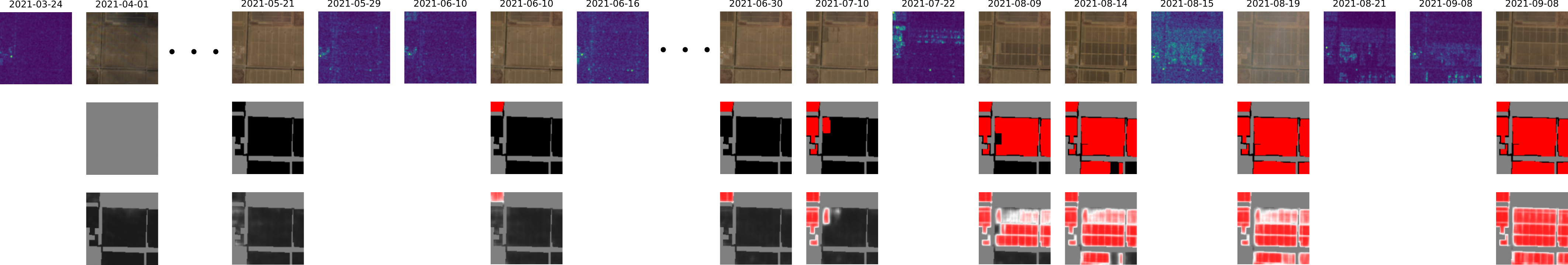}
    \caption{Multi-temporal solar panel segmentation with S2 and S1 SITS performed with the Time CosFormer. The top row corresponds to the input SITS. The second row corresponds to the Sentinel-2 labels, with black representing non-solar panel pixels and red representing solar panels. Grey is associated with unlabeled areas. The third row corresponds to the predictions of our proposed model after the sigmoid function. The colormap matches the label scheme (1 is associated with red, 0.5 with white, and 0 with black).}
    \label{fig:mm-sp}
\end{figure*}

\cref{fig:mm-sp} illustrates the input, labels and prediction in our multi-temporal solar panel segmentation tasks. We compare in \cref{tab:sp_results} the multi-temporal segmentation results obtained for solar panel detection mechanisms.

\begin{table*}[!htb]
\centering
\begin{tabular}{rlcccc}
\toprule
Variant & Model & Mod & Accuracy & F1 Score & IoU \\
\midrule
- & Transformer (non-causal) & S2 & $0.92 \pm 0.01$ & $0.84 \pm 0.02$ & $0.73 \pm 0.03$ \\
- & Transformer (causal) & S2 & $0.92 \pm 0.01$ & $0.85 \pm 0.01$ & $0.74 \pm 0.02$ \\
- & Linear & S2 & $0.92 \pm 0.01$ & $0.85 \pm 0.03$ & $0.73 \pm 0.04$ \\

- & CosFormer & S2 & $0.92 \pm 0.01$ & $0.85 \pm 0.02$ & $0.75 \pm 0.04$ \\
Time & CosFormer  & S2 & $0.92 \pm 0.01$ & $0.85 \pm 0.02$ & $0.74 \pm 0.03$ \\
- & LinRoFormer  & S2 & $0.91 \pm 0.01$ & $0.83 \pm 0.02$ & $0.71 \pm 0.02$ \\
Time & LinRoFormer  & S2 &$0.91 \pm 0.02$ & $0.84 \pm 0.04$ & $0.72 \pm 0.07$ \\
- & Retention & S2 &  $0.92 \pm 0.01$ & $0.85 \pm 0.02$ & $0.74 \pm 0.04$ \\
Time & Retention & S2 &$0.92 \pm 0.01$ & $0.84 \pm 0.03$ & $0.72 \pm 0.05$ \\
\midrule

- & Transformer (non-causal) & S1+S2 & $0.92 \pm 0.01$ & $0.85 \pm 0.02$ & $0.74 \pm 0.03$ \\
- & Transformer (causal)  & S1+S2 & $0.93 \pm 0.01$ & $0.86 \pm 0.02$ & $0.76 \pm 0.03$  \\
- & Linear & S1+S2 & $0.92 \pm 0.00$ & $0.85 \pm 0.01$ & $0.75 \pm 0.01$ \\

- & CosFormer &  S1+S2 & $0.93 \pm 0.01$ & $0.86 \pm 0.03$ & $0.76 \pm 0.05$ \\
Time & CosFormer & S1+S2 & $0.93 \pm 0.01$ & $0.86 \pm 0.01$ & $0.75 \pm 0.02$ \\
- & LinRoFormer &  S1+S2 & $0.91 \pm 0.01$ & $0.83 \pm 0.02$ & $0.72 \pm 0.03$ \\
Time & LinRoFormer & S1+S2 & $0.93 \pm 0.01$ & $0.86 \pm 0.02$ & $0.76 \pm 0.03$ \\
- & Retention & S1+S2 & $0.93 \pm 0.01$ & $0.86 \pm 0.02$ & $0.76 \pm 0.03$ \\
Time & Retention & S1+S2 & $0.93 \pm 0.01$ & $0.86 \pm 0.02$ & $0.76 \pm 0.03$ \\
\bottomrule
\end{tabular}
\caption{Multi-temporal solar panel segmentation metrics: mean and standard deviation across two different experiments. The \emph{Time} variant indicates modified reweighting mechanisms that account for acquisition dates.}
\label{tab:sp_results}
\end{table*}

We observe that all models trained with multi-modal data slightly achieve superior performance compared to their mono-modal versions, demonstrating the ability of these mechanisms to perform multi-modal fusion. In addition, for this task we also observe \df mechanisms matching transformer attention performance on both mono-modal and multi-modal settings. Among \df mechanisms, there are no major differences (performance gaps are often between 1-2 \% in F1 score or mIoU).

Nonetheless, LinRoFormer is the lowest-performing model on both tasks, with a gap of 4\% in mIoU in both mono-modal and  multi-modal settings compared to the best models. Still,  its temporal version, Time LinRoFormer, outperforms  it. This highlights the need to adapt existing mechanisms, to incorporate SITS specificities. Based on our previous experiments on look-back length influence, we hypothesize that this performance drop may be explained by different SITS lengths between training and testing.

\section{Conclusion}\label{sec:conclu} 
In this study, we have compared different \df mechanisms for multi-modal satellite image time series analysis. First, results obtained on both the supervised forecasting task and the multi-temporal segmentation task demonstrate the ability of these mechanisms to fuse multi-modal data. In particular, these mechanisms often match the performance of original causal attention \cite{Vaswani2017} while offering a recurrent formulation better suited for efficient inference. This result is promising as it opens new opportunities for regular and up-to-date land monitoring applications.
Second, we have studied the effect of the look-back length in both mono-modal and multi-modal tasks. For mono-modal forecasting, the results suggest that integrating temporal information via a temporal layer reweighting mechanism can slightly reduce forecasting error. Our experiments indicate that simple reweighting schemes, such as the CosFormer or our proposed Time CosFormer, achieve excellent performance compared to more complex mechanisms like the LinRoFormer or Retention. However, in the multi-modal setting, simple linear attention outperforms these reweighted dual mechanisms.

Furthermore, future works could extend our comparative analysis with other \df mechanisms such as State Space Models with Mamba 2 \cite{dao2024transformers} as well as measuring performances depending on the size of the recurrent state. 
Moreover, due to the scarcity of labeled data for numerous remote sensing applications, an important research direction is the development of self-supervised learning approaches for our proposed \df architecture. In this context, our proposed forecasting task could serve as an interesting starting point, as it is closely related to masked autoencoder strategies that have proven effective for pretraining SITS \cite{Dumeuralise2025,dumeur-2024-self-super,yuan22_sits_former}.

\section*{Acknowledgments}
This work was financed by the Agence Innovation Défense (AID), within the framework of the Dual Innovation Support Scheme (RAPID - Régime d'APpui à l'Innovation Duale), for the project 'DETEVENT' (Agreement No. 2024 29 0970).
This work was granted access to the HPC resources of IDRIS under the allocations 2025-AD011016513 and 2025-AD011012453R4 made by GENCI.
Sentinel-1 and Sentinel-2 data is from Copernicus. This paper contains modified Copernicus Sentinel data.

{
	\begin{spacing}{1}
		\normalsize
		\bibliography{local}
	\end{spacing}
}
\addvspace{30em}

\clearpage
\appendix

\section{ROPE} \label{ap:rotary}
The rotary matrix \cite{su2024roformer}, detailed in \cref{eq:rotary-matrix} operates $d/2$ $\theta_i$ rotations with $\theta_i=\frac{1}{10000^{2(i-1)/d}}$ with $i\in [1,.. d/2]$.

\begin{equation}\label{eq:rotary-matrix}
R_{t} = 
\begin{pmatrix}
\cos(t \theta_1) & -\sin(t \theta_1) & \cdots & 0 & 0\\
\sin(t \theta_1)& \cos(t \theta_1) &\cdots & 0  &0\\
\vdots  & \vdots  & \ddots & \vdots &\vdots  \\
0 & 0 & \cdots &  \cos(t \theta_{d/2}) & -\sin(t \theta_{d/2}) \\
0 & 0 & \cdots &\sin(t \theta_{d/2}) & \cos(t \theta_{d/2}) \\
\end{pmatrix}
\end{equation}
First,LinRoFormer applies a rotary matrix (noted $R_{t} \in \mathbb{R}^{d_k,d_k}$) on each query and key vectors, which will integrate relative positional encoding on the attention mechanism. 
\begin{equation}
    f(\mathbf{q},t)=\begin{pmatrix}
        q_1 \cos(t\theta_1) - q_2 \sin(t\theta_1) \\
        q_2 \cos(t \theta_1) + q_1 \sin(t \theta_1) \\
        \vdots\\
        q_{d-1} \cos(t \theta_{d/2}) -q_d \sin(t \theta_{d/2}) \\
        q_{d} \cos(t \theta_{d/2}) + q_{d-1} sin (t \theta_{d/2}) 
    \end{pmatrix}
\end{equation}

\section{Auxiliary variables encoding} \label{app:aux-encoding}
\subsection{AgERA5 variable} \label{app:agear5-encoding}
The forecaster takes as input a vector $\mathbf{w}'_{t+1}$ which corresponds to the encoding of the weather variables. In order to allow the model to learn relevant temoporal pattern, we use a cross-attention mechanism with learnable queries $Q_{\alpha} \in \mathbb{R}^{(n_q,d_{K_w})}$ and the input AgERA5 time series noted $W_{t+1}$. As indicated in \cref{eq:weather_encoder},  on each attention head, the input time series are projected to Keys and Values matrix using $W_{K_w} \in \mathbb{R}^{(n_w,d_{K_w})}$, $W_{V_w} \in \mathbb{R}^{(n_w,d_w)}$. The resulting weather vector correspond to a reshaped version of the projected time series $W'_t \in \mathbb{R}^{(n_q,d_w)}$ noted $\mathbf{w}'_t \in \mathbb{R}^{(n_q \times d_w)}$.
\begin{equation}\label{eq:weather_encoder}
W'_t=\text{Softmax}\left(\frac{Q_{\alpha} \cdot (W_t \cdot W_{K_w})^T}{d_K} \right)VW_{V_w}
\end{equation}
\subsection{S1 angles}\label{app:s1-angles}
When predicting a S1 acquisition, the forecaster takes as input the angle vector noted $\mathbf{a}^{S1}$. As detailed in \cref{eq:s1-angle}, this vector is an encoding of the azimuth $\phi$ and incidence angle $\theta$ of the S1 acquisition to predict. 
\begin{equation} \label{eq:s1-angle}
\mathbf{a}_t^{S1}=\text{Linear}\left([\cos(\phi_t);\sin({\phi_t});\cos(\theta_t)\sin(\theta_t)]\right)
\end{equation}

\subsection{Retention implementation}

To create a more consistent comparison to retention as implemented in the RetNet, we have operated several modifications. 
The queries and keys vectors are first encoded with $\phi(.)=1+ \text{elu}(.)$ and scaled $\frac{1}{d_k}$ to ensure more stability in the training.  

\end{document}